\documentclass[
    ,final            
  ]
  {aipproc}

\layoutstyle{6x9}

\begin{document}

\title{COULD SPIN-CHARGE SEPARATION BE THE SOURCE OF CONFINEMENT?}

\classification{11.15.-q, 12.38.Aw}
\keywords      {confinement}

\author{Antti J. Niemi}{
address={Department of Theoretical Physics,
Uppsala University 
P.O. Box 803, S-75108, Uppsala, Sweden; 
Laboratoire de Mathematiques et Physique Theorique
CNRS UMR 6083, Universite de Tours, ~~~
Parc de Grandmont, F37200, Tours, France}
}

\begin{abstract}
Yang-Mills gauge field with gauge group $SU(2)$ 
decomposes into a single charge neutral 
complex vector, and two spinless charged scalar fields. 
At high energies these constituents are tightly 
confined into each other by a compact $U(1)$ interaction, 
and the Yang-Mills Lagrangian describes the dynamics 
of asymptotically free massless gauge vectors.
But in a low energy and finite density 
environment the interaction between the constituents 
can become weak, and a spin-charge separation may 
occur. We suggest that the separation between the 
spin and charge with the ensuing condensation of the
charged scalars takes place when the Yang-Mills 
theory enters confinement. The confining
phase becomes then surprisingly similar to
the superconducting phase of a high-$T_c$ superconductor.
\end{abstract}

\maketitle


\section{Introduction}

According to popular folklore color (quark) 
confinement follows from an electric version 
of the BCS mechanism. This proposal is based on 
an assumption that the confining string is 
an electric version of the Abrikosov vortex \cite{Ri}

The Abrikosov vortex is present in a type-II superconductor,
where electrons condense into Cooper pairs.
It is a static string-like configuration along which
an undamped magnetic field line penetrates into the 
superconducting material. When such a magnetic vortex line 
forms between static particles with opposite magnetic 
charges (if such particles exist) it leads to a 
confining force that increases linearly in distance between 
the particles. 

But a magnetic vortex line does not lead to a 
confining force between static, electrically charged particles. 
For the confinement of electrically charged particles such 
as electrons, one needs vortex lines that conduct an 
electric field. However, the observation 
that magnetically charged point particles 
are confined by magnetic Abrikosov vortices provides an 
attractive picture for explaining the confinement of quarks:
Suppose the confining string is an analog of an Abrikosov vortex 
and suppose the quarks have a charge which couples 
to the component of the Yang-Mills field that is conducted
along the string. Then quark confinement can 
be explained in the same way as the confinement of 
magnetic point charges in type-II superconductors
is explained by (magnetic) Abrikosov vortices.

The quarks couple to the Yang-Mills field minimally, 
in the same manner as electrons couple to Maxwell's field 
in QED. As a consequence the confining string must couple to 
quarks in a manner which is different from the coupling between 
an electron and an Abrikosov vortex. Instead of a (nonabelian)
magnetic field, the confining string must be a carrier of
a (nonabelian) electric field, it must be an electric
dual version of the Abrikosov vortex. 

The BCS picture of quark confinement is consistent with the 
structure of N=2 and N=1 supersymmetric Yang-Mills 
theories \cite{seiberg}.
In these theories we have elementary Higgs fields that can describe the
Cooper pairing and condensation of magnetic monopoles.
This leads to an electric dual version of the Meissner effect 
and to the ensuing confinement of (nonabelian) electrically charged 
particles such as quarks.
This supersymmetry approach to confinement is intimately 
based on the existence and properties of the elementary Higgs fields,
and confinement is basically a consequence of a relatively
straightforward extension of the BCS theory. 

But in order to implement 
the BCS picture in a pure Yang-Mills
theory we first need to understand how to describe vortices in
an appropriate magnetic condensate.

In all known physical scenarios where vortices are present,
vorticity is supported by some kind of a medium. In ordinary
liquids such as helium
superfluids or water, a vortex is formed in a concrete
material environment. In a spontaneously broken (gauge) theory
vorticity is supported by a (material) condensation of the 
relevant order parameter. 

But in a pure Yang-Mills theory there is no apparent medium, no
elementary Higgs field that could condense.
Since there are no known vortex configurations 
that are formed in the absence of a supporting medium, we
have a fundamental problem in pure non-supersymmetric
Yang-Mills theory: The formation of a confining string between quarks
necessitates the introduction of a medium that carries vorticity. 
But there is no known mechanism how a medium could
be constructed or described in a pure Yang-Mills theory. 

In order to characterize a material environment that can support vorticity, 
we need some kind of a fundamental or effective (Higgs-like) 
field that can condense. In a pure Yang-Mills theory, the emergence of an 
effective Higgs field would mean that we can introduce 
some kind of a mechanism that leads to the formation of a 
condensation that consists of gluons. Since no such
gluonic version of Cooper pair formation is known, we then 
either need to develop new concepts and structures for 
describing vorticity, or 
alternatively we need to explain how an effective Higgs field 
could arise from outside of the pure Yang-Mills theory.

The Abrikosov vortex in a type-II superconductor is supported by a 
condensate that consists of Cooper pairs of electrons. As a consequence
it does not confine electrons, even though it can confine magnetically
charged point particles. Thus it is unlikely that the Cooper pairing of 
quarks can lead to a confining force between quarks. In order to explain 
quark confinement by a version of the BCS formalism, one needs instead
a Cooper pairing of (nonabelian) magnetically charged particles.
This means the confining flux tube must arise from the Yang-Mills 
field, and it receives no contribution from the condensation of quarks 
into (colored) Cooper pairs.

In the wider context of the Standard Model it is intrinsically
possible, but highly unlikely, that the Higgs field of the 
electro-weak sector could provide a condensate that also 
supports the confining string in the strong sector of the theory.
At the moment there are no theoretical arguments 
that anything like this could happen. The confinement of 
quarks appears to be an intrinsic property of the strong 
sector of the theory, with no contribution from the electroweak 
sector. Furthermore, at the moment we do not even have any experimental 
evidence that a fundamental electroweak Higgs exists. If it can 
not be found, we may well have a very similar problem in both 
the strong and electroweak sector of the standard model, the 
absence of a fundamental Higgs field that describes a condensate.

In a lattice formulation of Yang-Mills theory the problem of a
fundamental Higgs field can be avoided, by placing a singular vortex
line between the lattice sites. The finiteness of the lattice 
site then ensures the absence of singularities in the theory, 
at least as long as the lattice site is finite. But it remains 
to be explained how anything like this could be implemented in the 
continuum limit of the theory.

Finally, it could be that instead of a material vortex structure the
confining string has an intrinsic string theory description.
But in order to describe an intrinsic string, it is necessary to introduce
additional structures that are beyond a pure Yang-Mills theory: 
The intrinsic string approach to confinement would involve hypothetical
properties of the space-time that are at the moment unknown, 
besides that the pure Yang-Mills theory should emerge 
as a particular limit of 
the description.

Maybe 30 years of intense but unsuccesfull efforts by the 
theoretical community to construct a magnetic Cooper pair 
condensate in a pure 
Yang-Mills theory should be viewed as evidence that quark confinement 
can not be explained by the BCS formalism. 
In fact, we propose that there is no {\it a priori} 
reason why any version of the 
BCS formalism should explain confinement in a pure Yang-Mills 
theory, there is no evidence of any kind of magnetic Cooper 
pair formation. It could be that confinement in a pure, non-supersymmetric
Yang-Mills theory is due to an as yet unidentified 
mechanism which is quite different from the BCS picture.

Curiously, a very similar problem is also present in high temperature
superconductivity where the implementation of the BCS formalism
has thus far also failed: there is no theoretical or
experimental evidence that the electrons form Cooper pairs in 
superconducting cuprates \cite{wen}. 
While the Cooper pair formation
can not be definitely excluded, 
and there may even be some experimental support for a Cooper 
pair formation, the lack of any clear evidence for electron 
condensation into Cooper pairs has led to new ways 
for describing high-$T_c$ superconductivity. Curiously, the 
situation there is surprisingly 
similar to that in strong interaction physics:

\vskip 0.3cm

In the case of strong interaction physics, Yang-Mills theory
is widely accepted. Similarly, in the case of high temperature 
superconductivity there is a consensus that the materials can 
be described by a definite theory, the $t-J$ model.
In analogy to Yang-Mills theory, in this model there are no 
fundamental or effective Higgs fields that could support vortex 
structures with the ensuing Meissner effect. 
Consequently, at the moment, there is no theoretical understanding
how BCS formalism could be implemented to explain
high-T superconductivity. This has led to speculations that maybe
high temperature superconductivity is due to a mechanism which is
fundamentally different from the BCS formalism.

\vskip 0.3cm

Could it then be, that high temperature superconductivity in $t-J$
model has an origin which is similar to the origin
of quark confinement in a Yang-Mills theory?

\vskip 0.3cm

The lack of a Cooper pair in the $t-J$ model has led to
a very interesting theoretical proposal which, if correct, has far 
reaching consequences to our understanding of the fundamental 
structure of Matter. This proposal is based on the
very radical idea \cite{andersson}, 
\cite{wen} that in the strongly correlated environment of cuprate 
superconductors an electron ceases to be a fundamental particle. 
Instead an electron is a bound state of two 
other particles, which are called spinon and holon. The spinon 
is a fermion that carries the spin degree of freedom of 
the electron. It does not directly couple to Maxwell's 
electrodynamics. The holon is a spinless, complex boson 
and it carries the electric charge of the electron. Under 
normal circumstances the spinon and holon are tightly bound into
each other by a confining force, consistent with the
observational fact that at high energies an electron behaves
as a structureless point particle. But in the strongly correlated
environment of cuprate superconductors the force between the spinon
and holon could become weak, and a spin-charge 
separation may take place. A holon condensation can then provide
a material environment that support vorticity, leading
to the Meissner effect and an explanation of superconductivity \cite{wen}.

\section{Fermions}

In order to outline the slave-boson decomposition of an electron 
we start from a four-dimensional Dirac 
spinor $\psi^a_D$. Here $a=1,...,4$ label its four anticommuting 
components that obey the (graded) Poisson bracket 
\begin{equation}
\{ \psi^{a \dagger}_D (x) , \psi^b_D (y) \} = \delta^{ab} (x-y)
\label{0Dbr}
\end{equation}
We select the Weyl basis of the $\gamma$-matrices,
\[
\gamma^\mu = \left( \begin{array}{cc}
0 & \sigma^\mu \\ \bar \sigma^\mu & 0 \end{array} \right)
\]
where $ \sigma^0 = \bar \sigma^0$ is the $2\times 2$ unit matrix,
and $\sigma^i = - \bar \sigma^i
\ \ (i=1,2,3)$
are the standard Pauli matrices. In this basis we 
represent the Dirac fermion as
\[
\psi_D = \left( \begin{array} {c}
\xi_\alpha \\ \chi^{\dagger \dot \alpha} \end{array} \right)
\]
where $\xi_\alpha$
and $\chi^{\dagger\dot \alpha}$ (with $\alpha, \dot \alpha = 1,2$)
are two-component Weyl fermions. The spinor indices are 
raised and lowered using the
antisymmetric tensors $\varepsilon_{\alpha \beta}$ and
$\varepsilon^{\dot \alpha \dot \beta}$ with non-vanishing 
components determined by setting $ \varepsilon^{12} = 
\varepsilon_{21} = 1$. Explicitely, we have {\it e.g.}
$\xi^{\alpha} = \varepsilon^{\alpha \beta} \xi_\beta $
and $ \chi^{\dagger}_{ \dot \alpha } = \varepsilon_{\dot 
\alpha \dot \beta} \chi^{\dagger\dot \beta}$. 
Furthermore, when we introduce the conjugate variables 
$ \chi^\dagger_\alpha = (\sigma_0)_{\alpha \dot \beta} 
\chi^{\dagger \dot \beta}$ and $ \xi^{\dot \alpha} = 
(\bar \sigma_0)^{\dot \alpha \beta} \xi_\beta $
we get the graded Poisson brackets
\begin{equation}
\{ \, \chi^{\alpha} (x) , \chi^\dagger_\beta (y)\, \} 
= {\delta^\alpha}_\beta (x-y) \ \ \ \ \& \ \ \ \ \
\{ \, \xi^{\dagger}_{\dot \alpha} (x) , \xi^{\dot \beta} (y) \, \}
= {\delta_{\dot \alpha}}^{\dot \beta} (x-y)
\label{0bra}
\end{equation}
The relativistic version of the slave-boson decomposition 
is obtained by setting
\begin{equation}
\chi^\alpha = {\mathrm b}^\dagger \! \cdot \! {\mathnormal f}^\alpha 
+ \epsilon^{\alpha \gamma} {\mathnormal f}^{\dagger}_{\gamma} 
\! \cdot \! {\mathrm d}
\label{0wey}
\end{equation}
For the right-handed Weyl spinor  $\xi^{\dagger}_{\dot \alpha}$ we 
introduce an analogous decomposition, but here we do not need 
to display it explicitely. Here $\mathrm b$ and $\mathrm d$ 
are bosonic fields, they are the {\it holons} and  
subject to the Poisson brackets
\[
\{ \, \mathrm b^\dagger (x) , \mathrm b(y) \, \} = \{ \, \mathrm d^\dagger
(x) , \mathrm d (y) \, \}
= \delta(x-y)
\]
The $\mathnormal f_\alpha$ is an anticommuting (left-handed)
Weyl spinor. It is the {\it spinon} and it obeys the graded 
Poisson bracket
\[
\{ \mathnormal f^{\alpha} (x) , 
\mathnormal f^\dagger_\beta (y) \} \ = \ {\delta^\alpha}_\beta
(x-y)
\]
As a consequence, 
when we substitute the slave-boson decomposition (\ref{0wey}) in
(\ref{0bra}), we find that the decomposed Weyl fermion 
$\chi^\alpha$ obeys the graded Poisson bracket
\[
\{ \chi^{\alpha} (x) , \chi^\dagger_\beta (y) \} =
=
{\delta^\alpha}_\beta (x-y) \cdot \{ {\mathnormal f}^{\gamma}
{\mathnormal f}^\dagger_\gamma + {\mathrm b}^\dagger {\mathrm b}
+ {\mathrm d}^\dagger {\mathrm d} \}
\]
We also verify that
\[
\{ \chi^{\alpha} (x) , \chi^\beta (y) \} = 0
\]
Thus the decomposed field (\ref{0wey}) reproduces the
entire Poisson bracket structure of the original Weyl fermion 
$\chi^\alpha$ provided
we introduce the constraint
\begin{equation}
{\mathnormal N} =  {\mathnormal f}^{\gamma}
{\mathnormal f}^\dagger_\gamma + {\mathrm b}^\dagger {\mathrm b}
+ {\mathrm d}^\dagger {\mathrm d} = 1
\label{0con}
\end{equation}
With this constraint, the decomposition (\ref{0wey}) then becomes
an operator identity.

More generally, we can set 
\begin{equation}
{\mathnormal  N} =  {\mathnormal f}^{\gamma}
{\mathnormal f}^\dagger_\gamma + {\mathrm b}^\dagger {\mathrm b}
+ {\mathrm d}^\dagger {\mathrm d} = \mu
\label{0co2}
\end{equation}
where $\mu$ is some function. It can be selected arbitrarily, with 
the sole condition that $\mu(x)$ is non-vanishing for {\it all}
$x$. This ensures
that the resulting Poisson brackets of the decomposed fermion
$\chi^\alpha$ continue to define a graded symplectic two-form. 
The only difference between (\ref{0con}) and (\ref{0co2})
is, that when $\mu \not= 1$ the decomposed fermions are
graded canonical variables which are not of the Darboux form.

The condition (\ref{0con}), and its more general version (\ref{0co2}),
can be interpreted as the statement that for a separation between
spin and charge, the fermionic system must be in a physical environment
with a finite density, and the density is determined by the 
function $\mu(x)$. If this density vanishes for some $x$, the 
Poisson brackets of the decomposed fermion fail to reproduce 
the symplectic structure of the original fermion, and a spin-charge
separation can not occur. In particular, for all fields
${\mathnormal b} , \ {\mathnormal d} , \ {\mathnormal f}$
to have well defined Poisson brackets so that they
can be dynamical, each of the number densities 
${\mathnormal b}^\dagger {\mathnormal b}$, ${\mathnormal d}^\dagger 
{\mathnormal d}$, ${\mathnormal f}^\alpha {\mathnormal f}^\dagger_\alpha$
must be nonvanishing: An isolated electron can not become
decomposed into its spin and charge consituents, for a separation we need
a material finite density environment.

Both the holons $\mathrm b$ and  $ \mathrm d$ and the spinon
$\mathnormal f_\alpha$ are complex fields. Consequently the 
decomposition (\ref{0wey}) has an internal local $U(1)$ symmetry, 
the Weyl fermion $\chi^\alpha$ 
in (\ref{0wey}) remains intact when we send
\begin{equation}
\mathrm b \to  e^{i \theta} \mathrm b \ \ \ \ \ \& \ \ \ \ \
\mathrm d \to  e^{i\theta} \mathrm d \ \ \ \ \ \& \ \ \ \ \
\mathnormal f^\alpha  \to e^{i\theta} \mathnormal f^\alpha
\label{0sym}
\end{equation}
We note that this symmetry is generated by the 
canonical Poisson bracket action
of the number operator $\mathnormal N$ in (\ref{0con}). It is a 
{\it compact} $U(1)$ symmetry, that leads to an interaction between
the holons and spinons. For a large value of its coupling,
a compact $U(1)$ interaction is known to be confining. Thus
we expect that (\ref{0sym}) in general leads to an interaction 
between the spinons and holons which in a non-material environment 
where $\mu$ vanishes confines them into the (pointlike) fermion.

Conventionally, we couple Maxwell's eletromagnetism 
to the canonical charge operator defined by
\[
Q = \chi^\alpha \chi^\dagger_\alpha
\]
When we compute the canonical Poisson bracket
action of $Q$ on the Weyl spinor
$\chi^\alpha$ using the decomposed representation (\ref{0wey}), 
we get from (\ref{0co2})
\[
\{ Q (x) , \chi^\alpha (y) \} = {\mathnormal N}
(x) \cdot \chi^\alpha (x) \delta(x-y)
\ = \ \mu(x) \cdot \chi^\alpha(x) \delta(x-y)
\]
This states that $\mu(x)$ coincides with the local charge density
at $x$.
Clearly, this canonical action of $Q$ on the decomposed spinor 
can be reproduced by the canonical action of
\[
\bar Q = - \mu(x) \cdot [ \, {\mathrm b}^\dagger {\mathrm b} - 
{\mathrm d}^\dagger {\mathrm d} \, ]
\]
This confirms that the holons $\mathrm b$ and $\mathrm d$ become
(oppositely) charged under the standard coupling of a Weyl fermion
to Maxwellian electromagnetism,
while the spinon $\mathnormal f^\alpha$ is electrically neutral.
Thus the spinless holons indeed carry the entire electric 
charge of the Weyl (Dirac) fermion while its entire spin is 
carried by the charge neutral spinon.

In the ultraviolet, individual fermions such as quarks and 
leptons behave like structureless point particles. Consequently 
in the ultraviolet region there must be a very strong confining interaction
between their holon and spinon constituents. This is consistent with
the verity, that the $\beta$-function of an abelian gauge theory 
such as the compact $U(1)$ interaction between holons and 
spinons should not
display asymptotic freedom in the ultraviolet limit. 
Instead, it is natural to expect that the internal $U(1)$ interaction 
becomes strongly coupled and confining when we approach the 
ultraviolet limit. Thus the present slave-boson 
decomposition of a Dirac (Weyl) fermion is
consistent with the experimental 
observation that at high energies and low densities elementary 
particles such as leptons and quarks behave asymptotically
as structureless point particles. 

But at low energy scales it is feasible that 
a compact $U(1)$ theory becomes weakly coupled. 
In an infrared environment where the constraint (\ref{0con}) 
is obeyed, a Weyl fermion may then become split into its independent 
holon and spinon constituents. It has been proposed \cite{andersson}, 
\cite{wen}
that for an electron such a decomposition could take place in
strongly correlated cuprate superconductors. 
The ($d$-wave) high-T$_c$ superconductivity can then emerge in a 
phase where a spinon pairing becomes accompanied by a holon 
condensation,
\[
< b > \not= 0
\] 
with a consequential spontaneous breaking of the 
internal $U(1)$ symmetry. 

It is conceivable, that a slave-boson decomposition of a (relativistic)
fermion could
also occur in environments such as Early Universe when the density
was very large, or in the interior of hadronic matter when 
energies are not very high. In these high density environments 
the number operators for the holons and spinons are presumably 
nonvanishing
which implies that the ensuing Poisson brackets are nontrivial
so that both spinons and holons can become dynamical physical
degrees of freedom.

In order to test the relevance of the slave-boson decomposition 
in a given physical scenario, one needs in addition
to substitute the decomposed 
fermion into the corresponding Hamiltonian. One can then verify whether 
or not the spinons and holons can indeed describe propagating degrees 
of freedom in the environment of interest, in a normal manner. 
In the case of the $t-J$ model, under conditions that are
supposed to describe high-T$_c$ superconductivity, the decomposed
Hamiltonian does admit a natural intrepretation in terms of 
holons and spinons as particle-like excitations. This suggests,
that a separation between spin and charge may take place.
The theoretical and physical consequences of this scenario
have been discussed widely in the literature and we refer
to \cite{wen} for details.

\section{Gauge Fields}

We are curious, whether a similar separation between
spin and charge could also occur in the case of a Yang-Mills
theory, and whether this could lead to an understanding of 
confinement \cite{ludvig}, \cite{oma}, \cite{niels}. For simplicity
we shall only consider a pure $SU(2)$ Yang-Mills 
theory in a four dimensional space $\rm R^4$ with Euclidean 
signature. But a generalization to more general gauge group $SU(N)$ in
a Minkowskian signature space is straightforward.

We represent the gauge field
as a linear combination
\begin{equation}
A_\mu = A_{\mu i} \sigma^i = C_\mu \sigma^3 + X_{\mu +}\sigma^+ +
X_{\mu -} \sigma^-
\label{2A}
\end{equation}
where $\sigma^{\pm} = 1/2(\sigma^1 \pm i \sigma^2)$ and
\[
X_{\mu \pm} = A_{\mu 1} \mp i A_{\mu 2}
\]
\noindent
Our slave-boson decomposition of $A_\mu$ entails a decomposition 
of $X_{\mu \pm}$ into its spin and charge constituents. For this,
we introduce a complex vector field $e_\mu$ which we normalize 
according to
\begin{equation}
\begin{array}{ccc}
\vec e^2 & = & 0 \\
\vec e \cdot \vec e^* & = & 1
\end{array}
\label{2e}
\end{equation}
With $\psi_1$ and $\psi_2$ two complex scalars we can 
then write $X_{\mu\pm}$ as
\cite{ludvig}
\begin{equation}
X_{\mu +} = X_{\mu -}^* =
i \psi_1 e_\mu - \ i\psi^*_2 e^*_\mu
\label{2W+}
\end{equation}
Indeed, {\it any} four component complex vector  can 
always be represented as a linear combination of the
form (\ref{2W+}). For this, it suffices to observe
that an arbitrary, unconstrained four component
complex vector describes 
eight independent real field degrees of freedom. 
On the other hand, the two complex fields $\psi_{1}$ and $\psi_2$ 
describe four, and the complex vector $\vec e$ when subject to the
conditions (\ref{2e}) describes five independent
field degrees of freedom. But one of these corresponds
to the internal $U(1)$ rotation 
\begin{equation}
\begin{array} {ccc}
\vec e & \longrightarrow & e^{-i\xi} \vec e \\
\psi_1 & \longrightarrow & e^{i\xi} \psi_1 \\
\psi_2 & \longrightarrow & e^{i\xi} \psi_2 
\end{array}
\label{2int}
\end{equation}
which leaves the {\it r.h.s.} of (\ref{2W+}) intact.
As a consequence, in the general case the {\it r.h.s.}
of (\ref{2W+}) also describes eight independent field 
degrees of freedom. 

For simplicity, we may assume that the off-diagonal 
components $X_{\mu\pm}$ are subject to the maximal abelian
gauge condition
\begin{equation}
D_\mu^{ij }[C] X_{\mu j} = 
(\partial_\mu \mp i 
C_\mu) X_{\mu \pm} \ \buildrel {def} \over =
\ D^\pm_\mu X_{\mu \pm}
\label{2MAG}
\end{equation}
However, we shall not impose any condition on the diagonal 
component $C_\mu$. As a consequence the gauge condition
(\ref{2MAG}) removes two of the gauge degrees of freedom in 
$A_\mu$. This leaves us with a $U(1)\in SU(2)$
gauge invariance, which  corresponds to gauge transformations 
in the Cartan direction of $SU(2)$. Indeed, when we specify
\begin{equation}
g \to h = e^{i \omega \sigma^3}
\label{2h}
\end{equation}
we get
\begin{equation}
C_\mu \sigma^3 + X_{\mu +} \sigma^+ + X_{\mu -} \sigma^-
\ \buildrel {h} \over \longrightarrow \
(C_\mu + 2\partial_\mu \omega) \sigma^3 +
e^{2i\omega} X_{\mu +} \sigma^+ + e^{-2i\omega} X_{\mu -}
\sigma^-
\label{2hgt}
\end{equation}
while the condition (\ref{2MAG}) clearly remains intact.

When the $X_{\mu \pm}$ are 
subject to the condition
(\ref{2MAG}), in the representation (\ref{2W+}) there are
{\it a priori} restrictions both on the scalars $\psi_{1}$ and
$\psi_2$, and on the vector $\vec e$. 
But we now argue that (\ref{2MAG}) can be naturally interpreted as a 
restriction solely
on the absolute values $\rho_1$ and $\rho_2$ of the complex 
fields $\psi_{1}$ and $\psi_{2}$.
Indeed, consider the functional
\begin{equation}
\int d^4 \ x X_{\mu +} X_{\mu -} \ = \ \int
d^4\! x \, (|\psi_1|^2 + |\psi_2|^2)
\ = \ \int d^4 \! x ( \rho_1^2 + \rho_2^2)
\label{2val}
\end{equation}
This is manifestly invariant under the abelian gauge
transformation (\ref{2hgt}). But if we subject the {\it unconstrained}
$X_{\mu \pm}$ to an arbitrary infinitesimal $SU(2)$ gauge transformation 
and demand that (\ref{2val}) 
remains stationary, the ensuing Euler-Lagrange equation coincides
with the maximal abelian gauge condition \cite{zakh}
\[
\delta_g \! \int d^4 \! x X_{\mu +} X_{\mu -} = 0 \ \Rightarrow \
(\partial_\mu \mp i C_\mu) X_{\mu \pm} \ \equiv \ D^\pm_\mu X_{\mu\pm}
= 0
\]
Notice that the functional (\ref{2val}) involves only the two 
absolute values $\rho_1$ and $\rho_2$. Since 
the Euler-Lagrange equation {\it i.e.} the maximal abelian gauge
condition (\ref{2MAG}) gives 
two independent conditions, we can use it to solve for the two absolute
values $\rho_1$ and $\rho_2$ in terms of the other variables.
In the maximal abelian
gauge (\ref{2MAG}) both of the $\rho_{1}$ and $\rho_2$ 
then acquire their (gauge invariant) extrema values 
along the $SU(2)$ gauge orbit. 

We observe, that when we use the condition (\ref{2MAG}) 
and solve for $\rho_1$ and $\rho_2$, we introduce {\it no} restrictions 
on the complex vector $\vec e$. Nor do we 
introduce any restrictions 
on the phases of the complex fields $\psi_{1}$ and $\psi_2$. In particular,
this means that the internal symmetry (\ref{2int}) remains intact 
when we evaluate the absolute values $\rho_1$ and $\rho_2$ at their gauge 
invariant extrema along the gauge orbit.

We note that in general there are Gribov ambiguities in the
maximal abelian gauge condition. Consequently the extrema values
of $\rho_1$ and $\rho_2$ on the orbit are not unique. Here
we will not analyze the consequences that Gribov ambiguities might have.

The diagonal $U(1) \subset SU(2)$ gauge transformation (\ref{2hgt})
acts on the complex fields $\psi_{1,2}$ as follows,
\begin{equation}
\begin{array} {ccc}
\psi_1 & \to & e^{2i\omega} \psi_1 \\
\psi_2 & \to & e^{-2i\omega} \psi_2 
\end{array}
\label{2eu1}
\end{equation}
Here the phases differ from those in (\ref{2int}) by a relative
sign. Since this $U(1)$ transformation leaves the vector $\vec e$ intact,
only the complex fields $\psi_{1}$ and $\psi_2$  
couple to the Cartan subgroup $U(1) \subset SU(2)$.
On the other hand, the components $ e_\mu$
transform as a vector under Lorenz transformations
while the fields $\psi_{1}$ ad $\psi_2$ are scalars. This means
that (\ref{2W+}) entails a decomposition of 
$X_{\mu \pm}$ into two qualitatively very different
sets of fields: The scalar fields
$\psi_{1}$ and $\psi_2$ couple nontrivially to the 
abelian component of the $SU(2)$ gauge transformations {\it i.e.} 
carry a charge but have no spin.
The complex vector $\vec e$ is neutral {\it w.r.t.} the
abelian component of the gauge transformation but it
carries the spin degrees of freedom of the $X_{\mu\pm}$. 

As in the fermionic case,
for consistency of the decomposition (\ref{2W+})
we must assume that {\it both} condensates $\rho_{1,2}$ are
nontrivial. This means, that for a spin-charge decomposition
to occur in the quantum Yang-Mills theory we need {\it both} 
expectation values 
\begin{equation}
<\rho_{1,2}> = \Delta_{1,2} 
\label{2Del}
\end{equation}
to be nonvanishing. This condition then specifies the 
{\it material} environment where the separation between the spin and
the charge of a gauge field can occur. 

It is apparent that the present slave-boson 
decomposition of the gauge field is fully analogous to
the slave-boson decomposition of the Dirac (Weyl) fermion: In both
cases, the decomposition entails a separation between
the carriers of spin, and the carriers of charge. Furthermore,
in both cases the separation can only occur in a finite density
environment. In the case of a fermion we need the $\mu$ in
(\ref{0co2}) to be non-vanishing and in the case of gauge field
we need the condensates (\ref{2Del}) to be non-vanishing.
Furthermore, in both cases the decomposition introduces an internal,
compact $U(1)$ that can be employed to argue that asymptotically
in the short distance limit both the gauge field and the fermion
become structureless point particles, with the spinon and holon
confined to each other by the strong internal force. The internal spin-charge
structures can then be visible only in the infrared region and in
a finite density environment, when the internal $U(1)$ interaction
becomes weak. 

\vskip 0.5cm

In analogy with high-temperature superconductivity,
it becomes natural to propose that confinement
in $SU(2)$ Yang-Mills theory is described by a
a phase where spin-charge separation occurs 
and {\it both} condensates (\ref{2Del}) are 
nonvanishing, with the ensuing vortices describing the 
confining strings. There are tentative numerical results \cite{niels},
obtained by analysing the London limit of the Yang-Mills
quantum theory, that indicate that confinement can
indeed be related to the non-vanishing of
{\it both} order parameters (\ref{2Del}). But until now,
no serious lattice results have been presented to test 
this proposal. Such a serious lattice 
simulation would not only test whether 
the holon condensation could relate to confinement.
It would also test the fundamental structure of Matter, 
whether the known elementary particles could indeed be 
composites of more fundamental constituents that describe 
their independent spin and charge degrees of freedom.


\begin{theacknowledgments}
I thank Ludvig Faddeev for various discussions and 
collaboration that led to the development of the ideas
presented here. I also wish to thank the organizers of 
the $QCD@Work$ 2005 for giving me an opportunity to present my
results. I also wish to thank the Universities of Kyoto and Tokyo,
and APCTP for hospitality during this work. This research
has been supported by a VR Grant and by a STINT Thunberg scholarship. 
\end{theacknowledgments}


\end{document}